\documentclass[aps,prl,twocolumn,superscriptaddress]{revtex4}
\usepackage{graphicx}

\begin{document}

\title{Minority-spin $t_{2g}$ states and the degree of spin polarization in ferromagnetic metallic La$_{2-2x}$Sr$_{1+2x}$Mn$_2$O$_7$ ($x=0.38$)}

\author{Z. Sun}
\email{zsun@ustc.edu.cn}
\affiliation{Department of Physics, University of Colorado, Boulder, CO 80309, USA}
\affiliation{National Synchrotron Radiation Laboratory, University of Science and Technology of China, Hefei, Anhui 230029, P. R. China}

\author{Q. Wang}
\affiliation{Department of Physics, University of Colorado, Boulder, CO 80309, USA}

\author{J. F. Douglas}
\affiliation{Department of Physics, University of Colorado, Boulder, CO 80309, USA}

\author{H. Lin}
\affiliation{Department of Physics, Northeastern University, Boston, MA 02115, USA}

\author{S. Sahrakorpi}
\affiliation{Department of Physics, Northeastern University, Boston, MA 02115, USA}

\author{B. Barbiellini}
\affiliation{Department of Physics, Northeastern University, Boston, MA 02115, USA}

\author{R. S. Markiewicz}
\affiliation{Department of Physics, Northeastern University, Boston, MA 02115, USA}

\author{A. Bansil}
\affiliation{Department of Physics, Northeastern University, Boston, MA 02115, USA}

\author{A. V. Fedorov}
\affiliation{Advanced Light Source, Lawrence Berkeley National Laboratory, Berkeley, CA 94720, USA}

\author{E. Rotenberg}
\affiliation{Advanced Light Source, Lawrence Berkeley National Laboratory, Berkeley, CA 94720, USA}

\author{H. Zheng}
\affiliation{Materials Science Division, Argonne National Laboratory, Argonne, IL 60439, USA}

\author{J. F. Mitchell}
\affiliation{Materials Science Division, Argonne National Laboratory, Argonne, IL 60439, USA}

\author{D. S. Dessau}
\email{Dessau@colorado.edu}
\affiliation{Department of Physics, University of Colorado, Boulder, CO 80309, USA}

\begin{abstract}
Using angle-resolved photoemission spectroscopy (ARPES), we investigate the electronic band structure and Fermi surface of ferromagnetic La$_{2-2x}$Sr$_{1+2x}$Mn$_2$O$_7$ ($x=0.38$). Besides the expected two hole pockets and one electron pocket of majority-spin $e_g$ electrons, we show an extra electron pocket around the $\Gamma$ point. A comparison with first-principles spin-polarized band-structure calculations shows that the extra electron pocket arises from $t_{2g}$ electrons of minority-spin character, indicating this compound is not a complete half-metallic ferromagnet, with similar expectations for lightly-doped cubic manganites. However, our data suggest that a complete half-metallic state is likely to be reached as long as the bandwidth is mildly reduced. Moreover, the band-resolved capability of ARPES enables us to investigate the band structure effects on spin polarization for different experimental conditions.

\end{abstract}



\maketitle
Many materials such as manganites, Heusler metals and metallic oxides have been theoretically suggested to be half metals, in which the metallicity is completely dominated by electrons sharing one spin character, while electrons with the opposite spin are insulating, that is, electrons at the Fermi level are 100\% spin polarized \cite{deGroot,PickettPhysicsToday}. Rather than a theoretical toy, the concept of a half metal has vital applications in the field of spin electronics, in which the spins of electrons convey important information, in contrast to the traditional electronic devices where the charges of electrons process information \cite{Wolf}. In spite of extensive studies on many candidate materials, there is little definitive experimental evidence for half metallicity and experimental reports are often controversial \cite{NadgornyJPCM}. Moreover, when a ferromagnet is employed in spin electronic devices, its spin polarization may show strong variation with respect to different experimental circumstances depending upon whether the emphasis is on the density of states, ballistic transport, or diffusive transport properties \cite{NadgornyJPCM,Mazin}. This behavior arises from the detailed band structure of majority- and minority-spin states, which will put strong constraints on the applications of a ferromagnetic material \cite{NadgornyJPCM,Mazin}.

Manganites became the focus of attention recently, mainly because of the colossal magnetoresistive (CMR) effect in which the resistivity changes dramatically with an applied magnetic field \cite{Jin,Moritomo}. However, manganites are also an excellent candidate to be a half metal. In manganites, the spin of the mobile $e_g$ electrons interacts with the localized spins of $t_{2g}$ electrons via the Hund's rule. The mobile $e_g$ electrons have to avoid the strong on-site Hund's interaction $J$ $\sim$ 2.7 eV \cite{Dessauchapter}, which results in a gap for the minority-spin $e_g$ electrons. Therefore the splitting energy between the majority-spin and minority-spin bands is of the order of the Hund's rule $J$. When all $t_{2g}$ electrons align ferromagnetically, \textit{i.e.} the whole system is in the ferromagnetic state, the conducting electrons are expected to be highly spin polarized and therefore the system could be in a half metallic state. 

The best evidence of half metallicity in manganites was the spin-resolved photoemission study on cubic manganite La$_{0.7}$Sr$_{0.3}$MnO$_3$ by Park \textit{et al.} \cite{Park}, which directly observed 100\% spin polarization of the near-$E_F$ electronic states, implying a true half metallicity. However, this early data was taken from a thin film surface which did not show any angle dependence in the ARPES spectra, raising doubts about the nature of the surface.  Later data on a more well-controlled surface did not show full spin polarization, though it was interpreted as evidence for full spin polarization with a \textit{k}-perpendicular broadening from the 3-dimensional nature of the band structure in these materials \cite{Krempasky}. Such a true half-metallicity, if correct, has important implications for devices, for example those relying upon the Tunneling Magnetoresistive effect which has greatly increased contrast for full spin polarization \cite{NadgornyJPCM}. To date however, such TMR devices have not shown the high contrast expected from the half-metallicity promised by the spin-polarized photoemission experiments, instead implying polarization values from 54\% to 81\% \cite{Lu,JZSun}.  Further, point-contact tunneling and Andreev reflection experiments also indicated that La$_{0.7}$Sr$_{0.3}$MnO$_3$ is not a true half metal with extracted spin polarization values near 60\% for the cleanest samples and higher values for dirtier samples \cite{NadgornyJPCM,NadgornyPRB}.  Theoretically, LSDA band structure calculations do show evidence for minority spin states at the Fermi level \cite{PickettPRB}, though the addition of correlation effects in LSDA+\textit{U} calculations removes these states from the Fermi level for \textit{U} values $\sim$ 2 eV \cite{DessauPRL}, leaving the situation unclear. In addition, the study of minority-spin states is helpful for examination of theoretical models. For example, it has been proposed recently by Golosov that the presence of minority-spin states will introduce extra channels in microscopic electronic processes, which could be crucial for the physics of manganites \cite{Golosov}.

\begin{figure}
\includegraphics[scale=0.55]{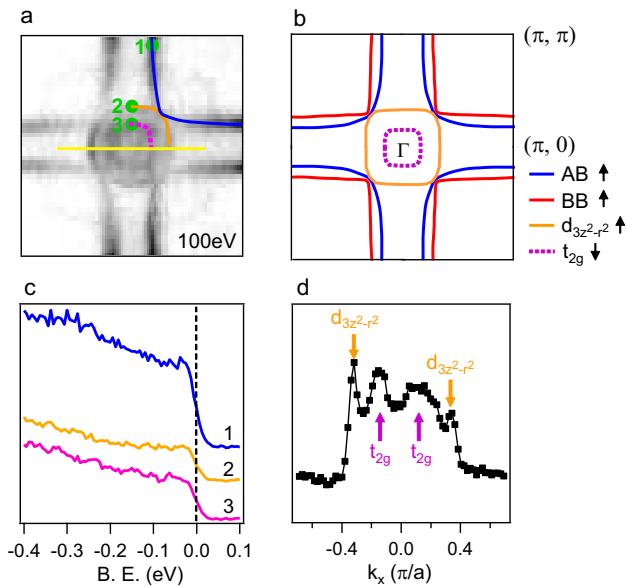}
\caption{(a) An experimental spectral weight distribution of La$_{2-2x}$Sr$_{1+2x}$Mn$_2$O$_7$ ($x=0.38$) taken at $E_F$ and T=30K using 100 eV photons. (b) A representative Fermi surface of La$_{2-2x}$Sr$_{1+2x}$Mn$_2$O$_7$ ($x=0.38$). These pockets come from bonding (BB), antibonding (AB), $d_{3z^2-r^2}$ and $t_{2g}$ bands. Majority and minority-spin characters of these Fermi surface segments are labeled as $\uparrow$ and $\downarrow$ respectively.  (c) Energy distribution curves taken at different locations (see panel a) in the first Brillouin zone. The different curves are vertically offset for clarity.  (d) A momentum distribution curve (MDC) at $E_F$ taken along the yellow cut of panel a, showing both sides of each of the two electron-like pockets.}
\label{FS}
\end{figure}

Bilayer manganite La$_{2-2x}$Sr$_{1+2x}$Mn$_2$O$_7$ shares many similar physics with the cubic manganites, including the colossal magnetoresistive effect and an LSDA band structure which becomes a half-metal in LSDA+\textit{U} calculations when \textit{U} $\sim$ 2 eV \cite{DessauPRL}. In contrast to the cubic manganites however, these materials cleave readily and have a nearly 2-D nature of the band structure, making them ideal for ARPES experiments. In combination with the first-principles band structure computations with emphasis of majority- and minority-spin states, we used ARPES to investigate the electronic band structure of $x=0.38$ bi-layer manganite La$_{2-2x}$Sr$_{1+2x}$Mn$_2$O$_7$, as this compound is devoid of the pseudogap and has strong and sharp spectral features at the Fermi level \cite{SunPRL}. By taking advantage of the photoemission matrix element effects, we have clearly resolved an additional band crossing the Fermi level that has not previously been seen, and band calculations indicate that this band originates from minority-spin $t_{2g}$ electrons. Therefore, we show that La$_{2-2x}$Sr$_{1+2x}$Mn$_2$O$_7$ ($x=0.38$) is not a complete half metallic ferromagnet, though we suggest that a mild tweak of the bandwidth may result in a true half-metal.  Moreover, using the parameters of mobile carriers extracted from individual bands directly probed by ARPES, we are able to study the band structure effects on the degree of spin polarization for different experimental circumstances, especially for electric transport properties.

The single crystals of La$_{2-2x}$Sr$_{1+2x}$Mn$_2$O$_7$ ($x=0.38$) were grown using the traveling-solvent floating zone method as described elsewhere \cite{MitchellPRB}. Our experiments were performed at beamlines 7.0.1 and 12.0.1 of the Advanced Light Source, Berkeley, using Scienta electron spectrometers. All samples were cleaved and measured at T$\sim$25-30K in a vacuum better than 3$\times$10$^{-11}$ Torr, and in-situ low energy electron diffraction (LEED) patterns confirmed the high quality of the surfaces. The combined instrumental energy resolution was better than 20 meV at beamline 12.0.1 and $\sim$ 30 meV at beamline 7.0.1, and the angular resolution was better than 0.025 $\pi/a$ and 0.04 $\pi/a$, respectively. First-principles band calculations of La$_{2-2x}$Sr$_{1+2x}$Mn$_2$O$_7$ ($x=0.5$) compound with ferromagnetic spin ordering were performed within the all-electron full-potential Korringa-Kohn-Rostoker (KKR) and linearized augmented plane-wave (LAPW) methods \cite{Bansil,Blaha}. The results are consistent with our previous studies \cite{Mijnarends}. The band structure for various dopings was obtained by adjusting the Fermi level to accommodate the correct count of electrons. Very similar results are obtained if the virtual-crystal approximation (VCA) is used to model the effects of La/Sr disorder \cite{HLin}.

Fig. 1(a) shows an experimental Fermi surface topology of bi-layer manganite La$_{2-2x}$Sr$_{1+2x}$Mn$_2$O$_7$ ($x=0.38$) taken with 100 eV photons, consisting of the antibonding hole pocket around the zone corners (the bonding hole pocket is invisible at this photon energy due to the matrix elements \cite{SunPRL}) and two electron pockets around the zone center $\Gamma$, which are schematically plotted in Fig. 1(b). As revealed by many theoretical and experimental studies, the two hole pockets arise from bonding and antibonding bands of bi-layer splitting band structure in this material with hybridized $d_{x^2-y^2}$ and $d_{3z^2-r^2}$ states of $e_g$ electrons, and the outer electron pocket consists mainly of $d_{3z^2-r^2}$ $e_g$ states \cite{DessauPRL,SunPRL,deJong,SunPRB,Saniz,Jozwiak}. However, the inner electron pocket has not yet been explicitly presented by ARPES investigations. As will be discussed later, this pocket arises from the minority-spin $t_{2g}$ states. Moreover, we have measured the Fermi surface using 130, 135, 140 and 145 eV (see the Supplementary Fig. S1), which are nearly photon energy (or $k_z$) independent, consistent with the quasi two-dimensionality of the electronic structures \cite{Saniz,Felser,deBoer,Huang}.

At low temperatures, the bonding and antibonding bands of the $x=0.38$ sample possess sharp Fermi cutoffs near the zone boundary \cite{SunPRL,deJong}, consistent with the low-temperature metallicity of this compound. In Fig. 1(c), we show the energy distribution curves (EDCs) taken from 3 points indicated in Fig. 1(a) to demonstrate the metallicity of antibonding band (EDC1), $d_{3z^2-r^2}$ states (EDC2), and the minority-spin $t_{2g}$ band (EDC3). It is evident that both the majority-spin $e_g$ bands and the minority-spin $t_{2g}$  bands show sharp Fermi cutoffs. This behavior is in contrast with doping levels of $x$ $\geq$ 0.4 which exhibits unusual coexistence of both metallic and pseudogap features in the EDCs \cite{DessauPRL,Chuang,Mannella}. The metallic nature of the minority-spin $t_{2g}$ states suggests that they participate in the electronic transport.  Contrary to ARPES data taken using lower photon energies \cite{SunPRL,SunPRB}, the quasiparticle peaks are invisible here, probably because the high photon energy and lower energy resolution are unfavorable for the detection of these small features. Though some reports suggest that the outmost layer of this material is probably insulating with A-type antiferromagnetism \cite{Freeland,Nascimento}, ARPES measurements consistently show that the metallicity of the bulk properties can be explicitly detected by performing experiments in ultrahigh vacuum better than 10$^{-10}$ torr \cite{SunPRL,deJong,SunPRB,Mannella}. In this study, our data clearly shows the bilayer-split band structure expected for ferromagnetic as opposed to antiferromagnetic ordering (see the Supplementary Fig. S1) as well as sharp Fermi edge cutoffs of the EDCs in Fig. 1(c) expected for a metallic as opposed to insulating state. This behavior is compelling evidence that our data are representative of the metallic properties of the bulk, regardless of the possible insulating top layer.

\begin{figure}[tbp]
\includegraphics[scale=0.6]{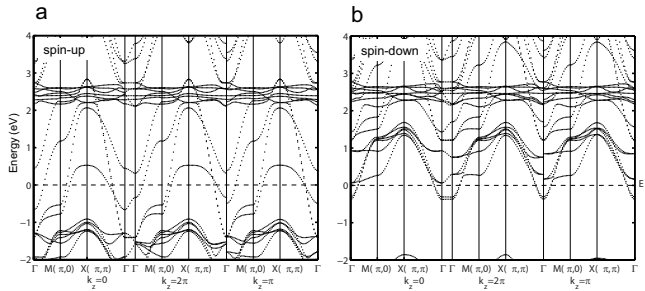}
\caption{Calculated band structure for the majority-spin states (a) and minority-spin states (b) of the $x=0.38$ compound. A small portion of the minority spin states crosses the Fermi energy near the $\Gamma$ point.}
\label{calculation}
\end{figure}

Band calculations for both majority-spin and minority-spin states have been performed. For the majority-spin states (Fig. 2(a)), itinerant $e_g$ electrons constitute the valence bands near the Fermi level and the $t_{2g}$ electrons form narrow bands at binding energy 1-2 eV. Because of the strong on-site Hund's interaction, most of the minority-spin $e_g$ and $t_{2g}$ bands are raised to higher energy above $E_F$ as shown in Fig 2(b). However, some minority-spin $t_{2g}$ electrons remain below $E_F$ around the zone-center. This result is consistent with previous band calculations, which suggest minority-spin states exist at the zone center and could be below the Fermi level \cite{SunPRB,Saniz,Felser,deBoer,Huang}. Comparing the experimental data with band structure calculations, we expect that the inner electron pocket around $\Gamma$ is derived from minority-spin $t_{2g}$ electrons. Although the band calculations suggest two $t_{2g}$ bands at the zone center, they are so close to each other that they cannot be decoupled in our experiments. We take into account this degeneracy when calculating the number of down-spin states.

In order to most clearly resolve the minority-spin $t_{2g}$ states near $E_F$, we tuned the ARPES matrix elements by changing photon energies and moving into other Brillouin zones. Fig. 3(a) shows a clear dispersive band of minority-spin $t_{2g}$ states with the bottom at $\sim$ 50meV, which was taken in the higher zone near ($2\pi/a$, $2\pi/a$) as shown in the inset. This dispersive band, with $k_F$ crossing of (0.15 $\pi$/a, 0.15$\pi$/a), constitutes the inner electron pocket in Fig. 1(a). Fig. 3(b) shows the energy-momentum distribution of $t_{2g}$ minority states and $d_{3z^2-r^2}$ majority states predicted by band structure calculations, with $k_z$ dispersion taken into account as a broadening effect. In order to match the experimental $k_F$, we elevate the calculated minority-spin $t_{2g}$ bands and obtain a dispersion (see the dotted line) with a bottom $\sim$ 100meV. Compared to the experimental band bottom (Fig. 3(a)), the $t_{2g}$ minority states are renormalized by a factor $\sim$ 2, and the resulting Fermi velocity is $\sim$ 0.55 eV\textperiodcentered\AA.  This knowledge of the Fermi velocity allows us to determine how the minority spin states contribute to the transport properties. 

 In bilayer manganites, the straight segments constitute most of the Fermi surface, which should dominate in transport properties.  Based on the experimental area enclosed by individual Fermi surface pockets, we estimate that the minority-spin states are only 4\% of total charge carriers, smaller than the theoretical prediction of 7\%. Similar estimated value can be deduced from magnetic Compton scattering experiments that have been performed on La$_{2-2x}$Sr$_{1+2x}$Mn$_2$O$_7$ ($x=0.4$)  at T = 5 K under high magnetic field of 7 T \cite{Yinwan}. The measured spin magnetic moment is $\sim$ 3.4 $\mu_B$. At $x=0.4$, the maximal value of magnetic moment given by the Hund's rule is 3.6 $\mu_B$, and one can estimate that the Fermi surface pocket contains about 0.2 electrons of minority-spin states, which gives a contribution of 0.2/3.6 = 5.6 \% of the total Fermi surface volume. This number is consistent with the value for $x=0.38$ estimated in the present paper. Moreover, the complex interactions in this material may make extra difference between theoretical and experimental values. Based on the volume enclosed by bonding, antibonding, $d_{3z^2-r^2}$ bands in our ARPES data, we estimate the spin polarization  $P_0 = \frac{N_\uparrow(E_F)-N_\downarrow(E_F)}{N_\uparrow(E_F)+N_\downarrow(E_F)}$
$\approx$ 92\%, with emphasis on density of states.

\begin{figure}[tbp]
\includegraphics[scale=0.5]{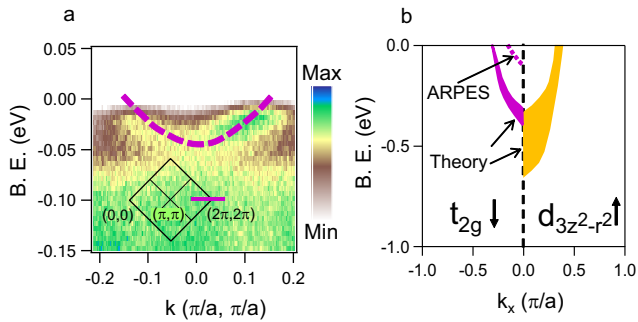}
\caption{(a) Experimental band structure at the zone center taken in the second zone (see the inset) using 56 eV photons. (b) The energy-momentum distribution of theoretical majority-spin $d_{3z^2-r^2}$ states (orange) and minority-spin $t_{2g}$  states (purple) with $k_z$ dispersion taken into account as a broadening effect. The dispersion of minority-spin $t_{2g}$ states (dashed purple line) as determined by ARPES. For best agreement with the ARPES we must shift the theoretical dispersion upwards (black arrow) to match the $k_F$ position, followed by a mass change of a factor of 2 to match the renormalized ARPES dispersion.}
\label{t2g}
\end{figure}

In the spintronic applications, La$_{2-2x}$Sr$_{1+2x}$Mn$_2$O$_7$ ($x=0.38$) can serve as a medium for electron conduction. As a matter of fact, the degree of spin polarization in transport processes plays a more important role than the $P_0$ with emphasis merely on the density of states. In principle, two scenarios, ballistic transport and Bloch-Boltzmann transport, should be considered when one evaluates the spin polarization in transport processes. As shown in Fig. 4, charge carriers behave very differently in the different transport regimes. Based on the dispersions of the majority spin bonding, antibonding and $d_{3z^2-r^2}$ majority states shown in Refs. 17 and in 24, we can estimate the Fermi velocities $v_F$ of those straight segments, which are 3.4 eV\textperiodcentered\AA, 2 eV\textperiodcentered\AA, and 4.8 eV\textperiodcentered\AA, respectively, significantly larger than that of the minority-spin states. Given the large Fermi surface volume of these majority-spin states, we conclude that most of the contribution to the electrical transport properties come from the majority-spin states.  More quantitatively, combining the Fermi velocities and the Fermi surface volume of all individual bands, we can obtain the spin polarization for ballistic transport, $P_1 = \frac{N_\uparrow(E_F)v_{F\uparrow}-N_\downarrow(E_F)v_{F\downarrow}}{N_\uparrow(E_F)v_{F\uparrow}+N_\downarrow(E_F)v_{F\downarrow}}$ $\approx$ 99\%, and for Bloch-Boltzmann transport, $P_2 = \frac{N_\uparrow(E_F)v_{F\uparrow}^2-N_\downarrow(E_F)v_{F\downarrow}^2}{N_\uparrow(E_F)v_{F\uparrow}^2+N_\downarrow(E_F)v_{F\downarrow}^2}$ $\approx$ 100\% \cite{NadgornyJPCM,Mazin}.

\begin{figure}[tbp]
\includegraphics[scale=0.5]{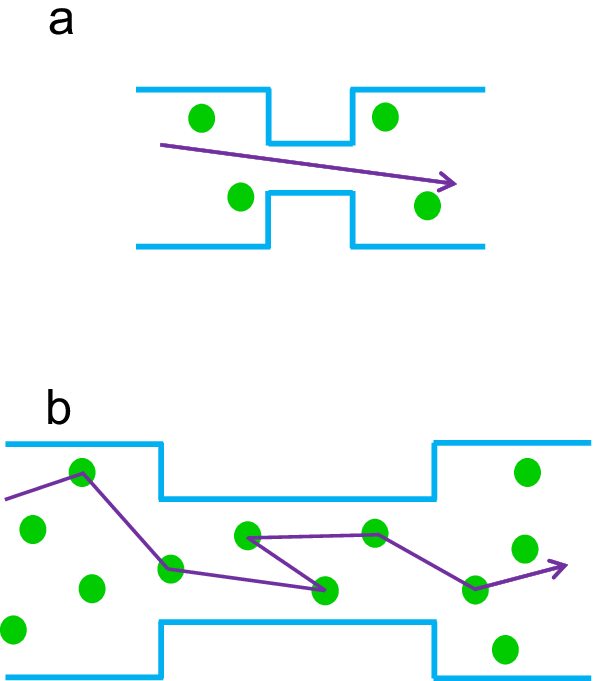}
\caption{(a) A sketch of ballistic transport regime. The mean free path of the mobile carriers are larger than the geometry size of the conductive medium. (b) A sketch of bulk Bloch-Boltzmann transport regime (or diffusive regime). The mean free path of the mobile carriers are smaller than the geometry size of the conductive medium.}
\label{fig4}
\end{figure}

The observation of minority-spin $t_{2g}$ states in La$_{2-2x}$Sr$_{1+2x}$Mn$_2$O$_7$ ($x=0.38$) suggests that in this material the conduction electrons are not fully spin-polarized at the Fermi level. In cubic manganite La$_{0.7}$Sr$_{0.3}$MnO$_{3}$, the bandwidth and the strength of the Hund's interaction (which is a local, on-site term) are similar to those in 2-D system La$_{2-2x}$Sr$_{1+2x}$Mn$_2$O$_7$. Therefore, the Hund's interaction probably cannot raise all minority-spin states of $t_{2g}$ electrons above the Fermi level, and the cubic manganite La$_{0.7}$Sr$_{0.3}$MnO$_{3}$ also is likely not a complete half metal. However, our results show that it may be promising to acquire a half metal of manganites by shifting the Fermi level down slightly (more hole doping), increasing the splitting between majority-spin $e_g$ bands and minority-spin $t_{2g}$ bands, or decreasing the band widths of minority-spin $t_{2g}$ states, which could eventually make all minority-spin states stay above $E_F$.  The easiest in principle is extra hole doping, which can be done with further Sr doping. However, in the bilayer manganite La$_{2-2x}$Sr$_{1+2x}$Mn$_2$O$_7$ further Sr doping just beyond the level we are at ($x \geq 0.4$), a spin-canted antiferromagnetic ground state appears, and the electronic structure changes drastically from $x=0.38$ \cite{DessauPRL,Chuang,Mannella}. This suggests that a simple doping procedure in La$_{2-2x}$Sr$_{1+2x}$Mn$_2$O$_7$ cannot eliminate the minimal minority-spin states and obtain a complete half metal.

In both the ballistic and Bloch-Boltzmann transports, the spin polarization is nearly 100\%. The calculations focus on the quasi two-dimensional electronic system of La$_{2-2x}$Sr$_{1+2x}$Mn$_2$O$_7$ because all of the transport properties are dominated by these in-plane states, which contain much higher Fermi velocities. Nevertheless, the large anisotropic resistivity suggests that the detailed spin transport may vary between the \textit{ab}-plane and \textit{c}-axis geometric setups. Our finding suggests that the metallic phase of La$_{2-2x}$Sr$_{1+2x}$Mn$_2$O$_7$ ($x=0.38$) behaves like a complete half metal in \textit{ab}-plane transport properties, which is largely due to the small Fermi surface volume and lower Fermi velocity of the minority-spin $t_{2g}$ states. However, our results indicate that the degree of spin-polarization can be manipulated by fabricating devices along different crystallographic directions, and to a smaller degree, with designed sizes so as to switch between the ballistic and Botzmann transport regimes.

In this paper, we have demonstrated that the combination of ARPES and ﬁrst-principles band calculations can indirectly reveal the spin characters of individual bands. Even for very tiny minority-spins states, this method is still applicable with high accuracy, and it will be very helpful for the studies of various half metals. In principle, as we have shown, the spin polarization of a material can be evaluated for various scenarios on the basis of band structure, which is valuable for designs of spintronic devices. Indeed, our finding suggests that a complete half-metal can be obtained in the transport behavior if the mobile carriers are confined to the Bloch-Boltzmann transport regime. However, the situation is more complicated when it comes to applications in real devices. For instance, the spin-flip scattering may occur at the interfaces, the mismatch of Fermi velocities across the interfaces introduces more complexities, and the anisotropy of individual $d_{x^2-y^2}$ and $d_{3z^2-r^2}$ bands may lead to different contributions for different geometric setups. More efforts are demanded to quantitatively take into account these facts to evaluate the applications of La$_{2-2x}$Sr$_{1+2x}$Mn$_2$O$_7$ ($x=0.38$) or similar materials, the transport properties of which mimic a complete half metal.

Primary support for this work was from the U.S. National Science Foundation under grant DMR 1007014, with supplementary support from the U.S. Department of Energy under grant DE-FG02-03ER46066.  The Advanced Light Source is supported by the Director, Office of Science, Office of Basic Energy Sciences, of the U.S. Department of Energy under Contract No. DE-AC02-05CH11231. The work at Northeastern University is supported by the US Department of Energy, Office of Science, Basic Energy Sciences contract number DE-FG02-07ER46352, and benefited from Northeastern University's Advanced Scientific Computation Center (ASCC), theory support at the Advanced Light Source, Berkeley and the allocation of time at the NERSC supercomputing center through DOE grant number DE-AC02-05CH11231. Z. S acknowledges the National Natural Science Foundation of China. Argonne National Laboratory, a U.S. Department of Energy Office of Science Laboratory, is operated under Contract No. DE-AC02-06CH11357.

\end{document}